\documentstyle{amsppt}
\Monograph
\loadbold

\NoRunningHeads
\catcode`\@=11
\def\logo@{}
\def\proclaimheadfont@{\smc}
\def\demoheadfont@{\smc}

   \catcode`\@=\active

    \magnification=\magstephalf

    \baselineskip 18pt


\LimitsOnSums
\LimitsOnNames
\CenteredTagsOnSplits

\def\addr#1{{\eightpoint\phantom{.}\hskip 7.5truecm#1 \newline}}

\def\theo#1{\procla{Theorem #1} }
\let\endtheo\endproclaim

\def\lem#1{\procla{Lemma #1} }
\let\endlem\endproclaim

\def\proof#1{\demol{Proof #1}}
\let\endproof\enddemo

\def\prop#1{\procla{Proposition #1}}
\let\endprop\endproclaim

\def\cor#1{\procla{Corollary #1}}
\let\endcor\endproclaim

\def\defi#1{\demol\nofrills{Definition #1.\usualspace}}
\let\enddefi\enddemo

\def\rem#1{\demol\nofrills{Remark #1.\usualspace}}
\def\rems#1{\demol\nofrills{Remarks #1.\usualspace}}
\let\endrem\enddemo

\let\demol\demo
\let\procla\proclaim

\define \bo {\bold}
\define \oti {\otimes}

\define\pc#1#2{ \frac {\partial {#1}}{\partial {#2}} }
\define\pcc#1#2{ \frac {\eth {#1}}{\eth {#2}} }
\define\pct#1#2#3{ \frac {\partial^{#1} {#2}}{\partial {#3}^{#1}} }
\define\pzt#1#2#3{ \frac {d^{#1} {#2}}{d{#3}^{#1}} }
\define\pz#1#2{ \frac {d{#1}}{d{#2}} }

\def\gd{\hbox{$\widehat{G}$}}

\def\kreuzprodukt{\hbox{$G\times_{\alpha} C_0(X)$}}
\def\nkreuzprodukt{\hbox{$G\times_{\alpha} L^{\infty}(X)$}}

\def\xd{\hbox{$\widehat{X}$}}
\def\gd{\hbox{$\widehat{G}$}}
\def\ad{\hbox{$\hat{\alpha}$}}
\def\dkreuzprodukt{\hbox{$\xd\times_{\hat{\alpha}} C_0(\gd)$}}
\def\ndkreuzprodukt{\hbox{$\xd\times_{\hat{\alpha}} L^\infty(\gd)$}}
\def\ww{\hbox{$\widehat{W}$}}

\def\ccgx{\hbox{$C_c(G\times X)$}}

\def\kt{\hbox{ $\Cal{K}$}}
\def\tr{\hbox{tr}}

\def\dk#1#2{\hbox{$(\,#1\,|\,#2\,)$}}
\def\skp#1#2{\hbox{$\langle\,#1\,|\,#2\,\rangle$}}

\def\lzg{\hbox{$L^2(G)$}}
\def\lzdg{\hbox{$L^2(\widehat{G})$}}
\def\lzx{\hbox{$L^2(X)$}}
\def\lzdx{\hbox{$L^2(\widehat{X})$}}
\def\lzgx{\hbox{$L^2(G\times X)$}}
\def\lzdgdx{\hbox{$L^2(\widehat{X}\times \widehat{G})$}}
\def\lzdgx{\hbox{$L^2(\widehat{G}\times X)$}}

\def\dg{\hbox{$\widehat{G}$}}

\parindent=0em

\topmatter
\title Random Operators and Crossed Products \endtitle
\author Daniel H. Lenz     \endauthor
\address {J. W. Goethe Universit\"at Frankfurt/Main}\endaddress
 \email {dlenz@math.uni-frankfurt.de} \endemail
\date January 1999      \enddate
{Random Operators}{ Dynamical Systems}{Density of states}  \endkeywords
\abstract 
This article is concerned with crossed products and their applications to random operators. We study the von Neumann algebra of a dynamical system using the underlying Hilbert algebra structure. This gives a particularly easy way to introduce a trace on this von Neumann algebra. We review several formulas for this trace, show how it comes  as an application of Connes' non commutative integration theory and discuss   Shubin's trace formula. We then restrict ourselves to the case of an action of a group on a group and include new proofs for some theorems of Bellissard and Testard on an analogue of the classical Plancherel Theorem. We show that the integrated density of states is a spectral measure in the periodic case, therby generalizing a result of Kaminker and Xia. Finally, we discuss duality results and apply a method of Gordon et al.  to establish a duality result for crossed products by   Z.
\endabstract

\endtopmatter

\document
\head{0. Introduction}\endhead

Families of random operators arise in the study of disordered media. More precisely, one is given a topological space $X$ and a family of operators $(H_x)_{x \in X}$ on $L^2(G)$.  Here, $X$ represents the set of "all   manifestations" of  a fixed kind of disorder on the locally compact abelian group $G$ \cite{3,4}.
 
The simplest example of a disordered medium is given by the periodic structure of a crystal. In this case $X$ is the quotient of $G$ by the subgroup of periods.  In the general case $X$ will not be a quotient of $G$, but there will still be an action $\alpha$ of $G$ on $X$. The fact that all points of $X$ stem from the same kind of disorder structure  is taken account of by requiring the action to be ergodic.

Whereas for a fixed $x \in X$ the operator $H_x$ may not have a large symmetry group, the whole family  of operators will be $G$-invariant.  This leads to the study of this family as a new object of interest.  This study is best perfomed in the context of $C^*$-algebras. In fact, it turns out that the crossed products $G\times_\alpha C(X)$ provide a natural framwork for these objects \cite{3,6,7,10,36}.

As it is, one is even led to a more general algebraic structure, viz $C^*$-algebras of groupoids when studying certain quasicrystals modelled by tilings \cite{21,26,27}.  But this is not considerd here.

In a remarkable series of papers Bellissard and his Co-workers introduced a K-theory based method called "gap labelling" for the study of random operators \cite{3,4,5,19}. Using results of Pimsner and Voiculescu \cite{32}, they were able to get a description of the possible gaps in many important cases.

As K-theory is best known in the cases, where either $G$ is discrete or $X$ stems from an almost periodic function,   much of their work was devoted to these cases. However, there are many important examples of more general random operators \cite{11,12,28,29}.

This is one of the starting points of this article. In fact, the main purpose of  Sections 1 and 2 is to study the framework of general random operators. This is done by means of Hilbert algebras. Sections  3 and 4  are then devoted to special results in the field of random operators. More precisely, this article is organized as follows.

In Section 1 we introduce crossed products, study two important representations and revise their basic theory. Special attention is paid to  the relationship between symmetry properties of random operators and direct integral decompositions. 

In Section 2 we use Hilbert algebras to the study of the von Neumann algebras and the $C^*$-algebras of the dynamical systems of Section 1.  We use them to introduce a trace on these von Neumann algebras. We show that this trace coincides with the trace introduced by Shubin for almost periodic operators \cite{36} and with the trace studied by Bellissard and others  for discrete $G$ \cite{3,4,5}. Moreover, we discuss how it is connected with Connes' non commutative integration theory \cite{13}.

In Section 3 we study the case that $X$ is a group itself. We study the relation between the two representations introduced in Section 1.  We  provide proofs for some theorems first announced in  \cite{6} and \cite{7} (cf. \cite{2} as well), whose proofs never seem  to have  appeared in print. Moreover, we revise the Bloch theory  for periodic operators from the algebaric point of view. This point of view has the advantage that the operators in questions are neither required to have pure point spectrum nor to have a kernel. We show that the integrated density of states  is a spectral measure in this case for purely algebraic reasons. This  generalizes a result of Kaminker and Xia \cite{25} and simplifies their proof.

Finally, in Section 4, we adapt a method developped by Gordon et al. \cite{22} for the study of the almost Mathieu operator to general  crossed products by {  Z}.  

\head{1. The $C^*$-algebra $G\times_\alpha C_0(X)$}\endhead
To every dynamical system $(G,\alpha,X)$ a $C^\ast$-algebra   can be constructed 
called  the crossed  product
of $G$ and $C_0(X)$ and denoted by  $\kreuzprodukt$.    If
$X$ consists of only one point, then $\kreuzprodukt$  is nothing but the
group $C^\ast$-algebra $C^\ast(G)$. We will be concerned with two
special representations of $\kreuzprodukt$. 
For further details on general crossed products we refer to \cite{31,38},  for  details on topological dynamics and crossed products see \cite{39,40,41}.
\subhead{1.1 Basic Definitions}\endsubhead
A dynamical system is a triple $(G,\alpha,X)$ consisting of
{\parindent=1cm
\item{-} a separable, metrizable, locally compact, abelian group G, whose Haar measure
  will be denoted by ds,
\item{-} a separable, metrizable space X,
\item{-} a continuous action $\alpha$ of $G$ on $X$,
\par}
Moreover we will need 
{\parindent=1cm\item{-} an $\alpha$-invariant measure  on $X$ with $supp\, m=X$
\par}
to define the representations considered below. We emphasize that this measure is not needed to define the crossed product. 

The group $G$ is acting on $\lzg:=L^2(G,ds)$ by 
$$T_t:\lzg\to \lzg, \;T_t
\xi(s):=\xi(s-t),\,s,t\in G, \xi\in \lzg$$ and on $\lzx:=L^2(X,dm)$ by
$$S_t:\lzx\to
\lzx,\;S_t\xi(x):=\xi(\alpha_{(-t)}(x)),\,t\in G,x\in X,\,\xi\in
\lzx. $$ Given a topological space $Y$, we denote by $C_c(Y)$ (  $C_0(Y)$,
$C_b(Y)$ resp.) the algebra of continuous functions with compact support
(vanishing at infinity, being bounded resp). Let $\|\cdot\|_\infty$ denote the
supremum norm on either of these algebras. The crossed product
$\kreuzprodukt$ is defined in  the following way:

Equipped with multiplication, involution and norm defined by
{\parindent=3cm
\item{} $a \ast b(t,x):=\int_G a(s,x) b(t-s,\alpha_{(-s)}(x))\,ds$,
\item{} $a^\ast(t,x):=\bar{a}(-t,\alpha_{(-t)}(x))$,
\item{} $\|a\|_1:=\int_G\|a(s,\cdot)\|_{\infty}\,ds$,
\item{} $a,b\in \ccgx, \,t\in G, x\in X,$
\par}
$\ccgx$ becomes a normed $*$-algebra. In general, this algebra
is neither complete nor a $C^*$-algebra. It is easy to see that
$$\|a\|:=sup\{ \|\rho(a)\|\,:\, \rho\; \hbox{ Hilbert space representation
  of} \,\ccgx\}$$
defines a $C^*$-seminorm on $\ccgx$. In fact $\|\cdot\|$ is a norm, as
can be seen by using the representations  $\pi_x,\,x\in X$, to be defined
below. The crossed product $\kreuzprodukt$ of the dynamical system $(G,\alpha,X)$ is the completion of
$\ccgx$ with respect to $\|\cdot\|$. It is immediate from these
definitions that every representation of $(\ccgx,\|\cdot\|_1)$ has a unique
continuous extension to a representation of $\kreuzprodukt$. We will be 
concerned with   two special representations and their direct integral
decompositions.
{\rem{1} The algebra ${\Cal A}:=(\ccgx,\ast, ^\ast,\skp{\cdot}{\cdot})$
where involution and convolution are defined as above and
$\skp{\cdot}{\cdot}$ is the inner product on the Hilbert space $\lzgx$ can easily be seen
to be a Hilbert algebra (cf. \cite{18}) i.e. to fulfil the following conditions:
{\parindent=3cm
\item{(i)} $\skp{a}{b}=\skp{b^\ast}{a^\ast}$ $a,b \in {\Cal A}$;
\item{(ii)} $\skp{a\ast b}{c}=\skp{b}{a^\ast\ast c}$ $a,b,c\in {\Cal A}$;
\item{(iii)} For $a\in {\Cal A} $ the mapping $b \mapsto a\ast b$ is
  continuous;
\item{(iv)} $\{ a\ast b : a,b\in {\Cal A}\}$ is dense in $\lzgx$.
\par}

In particular, the action of $\Cal A$ on itself from the left yields a
representation of $\Cal A$ on $\lzgx$,  which can be extended to a
representation of $\kreuzprodukt$ on $\lzgx$. These considerations will
be given a more precise form in Section 2 in order to study $\nkreuzprodukt$.
Now we prefer to introduce two representations, that allow a direct
integral decomposition.
\endrem 
}
\subhead {1.2 Representations of $\kreuzprodukt$}\endsubhead

Let $\pi:\ccgx \longrightarrow  B(\lzgx)$ be given by 
$$ \pi(a) \xi(t,x):=\int_G a(t-s,\alpha_t(x)) \xi(s,x)\,ds,\; \xi\in \lzgx,$$
and let for $x\in X$ the representation  $\pi_x:\ccgx\longrightarrow B(\lzg)$ be given by
$$ \pi_x(a) \xi(t):=\int_G a(t-s,\alpha_t(x)) \xi(s)\,ds,\; \xi \in \lzg,$$
where $B(H)$ denotes the algebra of bounded operators on the
Hilbert space $H$. Then it is easy to see that $\pi$ and $\pi_x$, $x\in
X$, are continuous representations of $\ccgx$. Their  extensions  to $\kreuzprodukt$
will also  be denoted by  $\pi$ and $\pi_x$. 
Identifying $\lzgx$ with $\int_X^\oplus \lzg \, dm$, we get
$$\pi(A)=\int_X^\oplus \pi_x(A)\,dm,\;\,A\in\kreuzprodukt.$$
This is obvious for $A\in \ccgx$ and follows for arbitrary $A\in
\kreuzprodukt$ by  density. As $G$ is amenable, even abelian, the
representation $\pi$ is faithful (cf. §7.7 in \cite{31}). Therefore we
have
$$\kreuzprodukt\simeq \pi(\kreuzprodukt)=\overline{\pi(\ccgx)}.$$
Thus the crossed product is the norm closure of an algebra of certain 
integral operators. 

We remark that the mapping $X\ni x\mapsto \pi_x(A)$
is strongly continuous for $A\in \kreuzprodukt$, as can be directly
calculated for $A\in \ccgx$, and then follows by density arguments for all
$A\in \kreuzprodukt$. The representation $\pi$ has two interesting
symmetry properties, that will be given in the next proposition.
\prop{1.2.1}
{(a)} Let $\Cal{D} $ be  the algebra of diagonalisable operators
  on the Hilbert space $\lzgx \simeq \int_X^\oplus \lzg \, dm $, then
$$\pi(\kreuzprodukt)\subset D'.$$

{(b)} $\pi(\kreuzprodukt)\subset\{T_t\otimes S_{-t}\,:\,t\in
  G\}'$. Moreover for   $t\in G,\,x\in X$ and $A\in \kreuzprodukt$ the formula
 $T_t\, \pi_{\alpha_t(x)}(A)\, T_t^\ast =\pi_x(A)$ holds.
\endprop
\proof{} (a) This is  the fact that $\pi(A)$ permits a direct integral decomposition.

(b) This can be directly calculated for $A\in \ccgx$ and then follows for arbitrary $A\in \kreuzprodukt $ by a density argument. \hfill $\qed$ \endproof


Moreover, the following is valid.

\prop{1.2.2.} If $\alpha$ is minimal i.e. $Gx:=\{\alpha_t(x):t\in
  G\} $ is dense in $X$ for every $x\in X$ then
{\parindent=1cm
\item{(i)}  $\sigma(\pi_x(A))$ is independent of $x\in X$ for
  selfadjoint $A\in \kreuzprodukt$,
\item{(ii)} $\pi_x$ is faithful for every $x\in X$.
\par}
If $G$ acts ergodically on $X$, then there exists for each selfadjoint $A\in  \kreuzprodukt$ a closed set $\Sigma_A\subset R$ and a measurable set $X_A\subset X$ with ´$\mu(X-X_A)=0$ s.t. $\sigma(A_x)=\Sigma_A$ for all $x\in X_A$.
\endprop
\proof{} The first statement is proven in \cite{25}, the second one in Section 4 of \cite{3}. \hfill $\qed$ \endproof

We will now give a second representation of $\kreuzprodukt$. For
${\hat t}\in \hat{G}$ let $\pi^{{\hat t}}:\ccgx\rightarrow B(\lzx)$ be
defined by
$$\pi^{{\hat t}}\xi(x):=\int_G
a(s,x)\xi(\alpha_{-s}(x))\dk{\hat t}{-s} ds,\;a\in \ccgx,\;\xi\in \lzx,$$
where $\dk{\cdot}{\cdot} $ denotes the dual pairing between $G$ and
$\widehat{G}$.  For  ${\hat t}\in \widehat{G}$ the mapping  $\pi^{{\hat t}}$ is then  a representation
of $\ccgx$, which has a 
unique continuous   extension to a representation of $\kreuzprodukt$, 
again  denoted by $\pi^{{\hat t}}$.  For $A\in \kreuzprodukt$ the
mapping ${\hat t}\mapsto \pi^{{\hat t}}(A)$ is strongly continuous, as can
be seen using the same arguments as in the case of the mapping $x\mapsto
\pi_x(A)$. Therefore, we can define a  representation
$$\widetilde{\pi}:=\int_{\widehat{G}}^\oplus\pi^{{\hat t}}\,d
{\hat t}:\kreuzprodukt\rightarrow B(\,\int_{\widehat{G}}^\oplus \lzx d{\hat t}\, ),$$
where we denote by $d{\hat t}$ the Haar  measure on $\widehat{G}$.
Let the unitaries $W$ and $U$ be defined by
$$W:\lzgx\longrightarrow \lzgx, \;W\xi(t,x):=\xi(t,\alpha_t(x)),\;\xi
\in \lzgx,$$ 
$$U:=(F_G\otimes I) W^\ast\,:\lzgx\longrightarrow \lzdgx,$$
where $F_G:\lzg\longrightarrow \lzdg$ is the Fouriertransform and
$I$ the identity. Then
we have
$$\widetilde{\pi}=U\, \pi\, U^\ast,$$
where $\lzdgx$ is identified with $\int_{\widehat{G}}^\oplus \lzx
d{\hat t}. $

\rems{1} The crossed product $\kreuzprodukt$
is just the group $C^\ast$-algebra $C^\ast(G)$, if $X$ consists of only one point. In this case we  identify
$\ccgx$ with $C_c(G)$ and  we get 
$\pi(\varphi)\xi= \varphi\ast \xi=T_\varphi \xi,\; \,\widetilde{\pi}(\varphi)\xi = M(F(\varphi)) \xi,$
where  $T_\varphi$ denotes the operator of convolution with $\varphi \in
C_c(G)$ and $M(\psi)$ denotes the operator of multiplication with
$\psi$. This implies 
$\widetilde{\pi}(\kreuzprodukt)=M(C_0(\widehat{G}))$ and $ \pi(\kreuzprodukt)=\{\,F^{-1} M_\psi F\,|\, \psi \in C_0(\widehat{G})\}.$

2. The direct integral decomposition of $\widetilde{\pi}$ relies
essentially on the symmetry
$$\pi(\kreuzprodukt)\subset\{T_t\otimes S_{-t}\,:\, t\in G\}',$$
as can be seen in the following way:
Using $W\, (T_t\otimes I)\, W^\ast= T_t\otimes S_{-t}$, one gets immediately
$ U (T_t\otimes S_{-t}) U^\ast =M_{t}\otimes I,$
where $M_t$ denotes the operator of multiplication with $\dk{t}{\cdot}$
on $\lzdg$. Therefore, we have
$ U\, \pi(A)\, U^\ast\in \{M_t\otimes I \,|\, t\in G\}', $
and this implies (cf. 5, ch 2, II in    \cite{18}) that $U \pi(A) U^\ast$ has
a direct integral decomposition. 
\endrem

\head {2. The von Neumann algebra $\nkreuzprodukt$}\endhead
In this section we study the von Neumann algebra $\nkreuzprodukt$.
We will be particularly interested in determining its generators and its commutant as well as introducing and calculating a trace on it.  In a sense, much more general considerations can be found in \cite{23}, where arbitrary von Neumann crossed products are studied by means of Tomita Takesaki theory of left Hilbert algebras (cf. \cite{38}). The trace on $\nkreuzprodukt$
 allows to introduce for each selfadjoint operator affiliated to
 $\nkreuzprodukt$ a canonical spectral measure. This spectral measure is
 called the {\it density of states}. We will discuss the so called {\it Shubins's trace formula}, relating the density of states
 to the number of eigenvalues of certain restricted operators. 
We conclude the section with the discussion of certain formulas for the
trace in the case that $m(X)<\infty$ holds.

\subhead{ 2.1  Definition and important properties of $\nkreuzprodukt$}\endsubhead

Following 7.10.1 in \cite{31} we  define the von Neumann crossed product.

\defi{2.1.1} $\nkreuzprodukt:=\pi(\kreuzprodukt)''  $ \enddefi

We will study this algebra by means of the already defined  Hilbert algebra ${\Cal A}=(\ccgx,\ast, ^\ast,\skp{\cdot}{\cdot})$ (cf. Remark 1 in Section 1.1). We need some notation.
\def\LA{\hbox{${\Cal L}({\Cal A})$}}
\defi{2.1.2}(a) For $A\in {\Cal A}$ let $L_a$ ($R_a$)  be the unique continuous operator with
$$ L_a \xi=a\ast \xi,\;\, (R_a \xi=\xi \ast a),\;\,  \xi \in \ccgx,$$
i.e. $L_a \xi=\int_G a(s,x)\xi(t-s,\alpha_{-s}(x)) ds$ for $\xi \in \lzgx$ and similarly for $R_a$.

(b) The unique  extension of $^\ast: \ccgx\longrightarrow \ccgx$  to a continuous mapping of   $\lzgx$ into itself will also be denoted by $^\ast$, i.e. $a^\ast(t,x)=\overline{a}(-t,\alpha_{-t}(x))$ for $a\in \lzgx$.

(c) An $a\in \lzgx$ is called left bounded (right bounded) if the mapping $\xi \mapsto R_\xi a$ ($\xi \mapsto L_\xi a$) can be extended to a continuous operator on $\lzgx$. This operator will be denoted be $L_a$ ($R_a$).

(d) ${\Cal L}({\Cal A}):=\{ L_a\;:\; a\in {\Cal A}\}''$, ${\Cal R}({\Cal A}):=\{ R_a\;:\; a\in {\Cal A}\}''$.
\enddefi

The  connection between these crossed products and Hilbert algebras is simple.
\lem{2.1.3} For $a\in \ccgx$ the equality
$ W^\ast \pi(a) W = L_a$
holds.
\endlem
\proof{} For $\xi \in \ccgx$  a direct computation yields $ W^\ast \pi(a) W\xi = L_a \xi$ and the lemma follows, as $L_a$ and $ W^\ast \pi(a) W$ are bounded and $\ccgx$ is dense in $\lzgx$. \hfill $\qed$
\endproof

The lemma and the definitions of $\nkreuzprodukt$ and $\LA$ directly yield

\theo{2.1.4} $Ad_W:\nkreuzprodukt\longrightarrow\LA$, $A\mapsto W^\ast A W$ is a spatial isomorphism of von Neumann algebras.
\endtheo

Those operators which are inverse images of left bounded operators under $Ad_W$ will play an important role.

\defi{2.1.5}
(a) A function $a\in \lzgx$ is called the kernel of the operator $A\in \nkreuzprodukt$ if $a$ is left bounded and $A=W L_a W^\ast$.

(b) Let $\kt:=\{A\in \nkreuzprodukt\,|\, \hbox{ A has a kernel}\}.$
\enddefi
We study $\kt$ in the next proposition.

\prop{2.1.6} (a) The operator $A$ has the kernel $a\in \lzgx$ iff
$$ A\xi(t,x)=\int_G a(t-s,\alpha_t(x)) \xi(s,x)\,ds\;\,\hbox{  a.e.}$$
holds for $\xi \in \lzgx$.

(b) The set $\kt$ is an ideal in  $\nkreuzprodukt$. For $A\in
  \kt$  with kernel $a\in \lzgx$ and $B\in \nkreuzprodukt$ the kernel
   of $AB$ is given by
$  W^\ast B (W a)$
and the kernel of $A^\ast $ is given by $a^\ast$.

(c) For  $A\in \nkreuzprodukt$ with kernel $a\in \lzgx$ the
  operator $A_x$ is a bounded Carleman operator with kernel
$$ a_x(t,s):=\overline{a(t-s,\alpha_t(x))}$$
(i.e. $(A_x)f(t)=\skp{a_x(t, \cdot)}{f}$ a.e. $ t\in G$) for a.e. $x\in
X$. 
\endprop

\proof{} (a) The statement with "$\xi \in \lzgx$" replaced by   "$\xi \in \ccgx$" is easy to calculate. Using that the maximal operator given by the integral expression is closed, we get (a).

(b) As $\kt$ contains the algebra    $\pi(\ccgx)$ by Lemma 2.1.3., it is strongly dense in  $\nkreuzprodukt$. The remaining statements could be calculated directly but also follow from Proposition 2 and Proposition 3 in 3, ch. 5, I of \cite{18}.

(c) We set $a_x(t,s):=\overline{a(t-s,\alpha_t(x))}$ for $a\in
\lzgx$. 
 Using the Fubini theorem it is easy to see that $a_x$ is the kernel
of a Carleman operator (cf. \cite{42}) $\widetilde{A_x}$ for almost every $x\in X$. It
remains to show $\widetilde{A_x}=A_x$, a.e. $x\in X$. For $\eta \in \lzx$
and $\xi\in \lzg$ a short calculation yields
$$ \eta(x) A_x \xi (t)= \eta(x) \widetilde{A_x}\xi(t), a.e.$$
As $\eta\in \lzg$ was arbitrary this implies
$$ A_x\xi=\widetilde{A_x}\xi, a.e.\; x\in X.$$
Using a countable, dense subset of $\xi \in \lzg$ and the fact that
$\widetilde{A_x}$ is closed, we conclude (c). \hfill $\qed$
\endproof
 
We can now characterize $\nkreuzprodukt$ and its commutant.

\theo{2.1.7} (a) $\nkreuzprodukt=W \LA W^\ast=\{T_t\otimes I,\; W (I \otimes M_v) W^\ast :\;\, \;t\in G, v\in L^{\infty}(X)\}''$.

(b)  $G\times_{\tilde{\alpha}} L^\infty(X)={\Cal R}({\Cal A})=\{T_t\otimes I,\, W^\ast (I\otimes M_v) W \,: \;t\in G, v\in L^{\infty}(X)\}''$ with $\tilde{\alpha}:G\times X\rightarrow X$, $\tilde{\alpha}_t(x):=\alpha_{-t}(x)$.

(c) $(\nkreuzprodukt)'=W {\Cal R}({\Cal A})   W^\ast=\{T_t\otimes S_{-t},\, I\otimes M_v\;:\,t\in G, v\in L^{\infty}(X)\}''$  
\endtheo
\proof{}(a) The equality  $\nkreuzprodukt=W \LA W^\ast$ has already been  pro\-ven in Theorem 2.1.4. To prove the second equality we set
$${\Cal C}:=\{T_t\otimes I,\; W (I\otimes M_v) W^\ast \,: \;t\in G, v\in L^{\infty}(X)\}.$$

${\Cal C}''\subset W \LA W^\ast$: By $\LA =  {\Cal R}({\Cal A})'$ (cf. Theorem 1 in 2,ch 5, I of \cite{18}), it is enough to show 
${\Cal C}\subset ( W  {\Cal R}({\Cal A}) W^\ast)'$, i.e.
$$ C W R_a W^\ast = W R_a W^\ast C,\;a\in \ccgx, C\in {\Cal C}.$$
This can be  calculated directly.

$W \LA W^\ast\subset {\Cal C}''$: For $u\in C_c(G)$ and $v\in C_c(X)$ and $\xi \in \lzgx$ an easy calculation yields
$$ \pi(u\otimes v) \xi(t,x)= (W (I \otimes M_v) W^\ast \xi(t,x)) \cdot \int_G u(s) (I\otimes T_s)\xi(t,x) ds,$$
and this implies
$\pi(u\otimes v)\subset {\Cal C}''.$ The desired inclusion follows.

(b) Defining $\pi_{\tilde{\alpha}}$ by simply replacing $\alpha$ by $\tilde{\alpha}$ in the definition of $\pi$, we get
$$R_a \xi = \pi_{\tilde{\alpha}}(Wa)\xi,\; a\in \ccgx,\xi\in \lzgx.$$
This implies
$ {\Cal R}({\Cal A}) =\{\pi_{\tilde{\alpha}}(Wa)\,:\,a\in \ccgx\}''=\{\pi_{\tilde{\alpha}}(a)\,:\,a\in \ccgx\}''.$

As $ G\times_{\tilde{\alpha}} L^\infty(X)=\{\pi_{\tilde{\alpha}}(a)\,:\,a\in \ccgx\}''$ by definition of the crossed product the first equality is proven.
The second equality follows by replacing $\alpha$ by $\tilde{\alpha}$ in  (a), i.e. by replacing $W$ by $W^\ast$.

(c) The first equality follows from (a) and ${\Cal R}({\Cal A})= \LA'$. The second equality follows by (b) and $W (T_t\otimes I) W^\ast = T_t\otimes S_{-t}$.
\hfill $\qed$
\endproof
Theorem 2.1.7 yields 
$\nkreuzprodukt\subset D'.$
In particular (cf. 5, ch,2, II in \cite{18}), every $A\in \nkreuzprodukt$ can be
written as a direct integral
 $A=\int_X^\oplus A_x \,dm$, whose fibres are uniquely determined up to a
 set of measure zero. Similarly it can be seen that for $A\in
 \nkreuzprodukt$ the equation
$$ U A U^\ast=\int_{\widehat{G}}^\oplus A^{{\hat t}} d {\hat t}$$
holds, where the $ A^{{\hat t}}$ are uniquely determined up to a set of
measure zero.
 From now on we will  identify $\kreuzprodukt$ with
$\pi(\kreuzprodukt)$. For $A$ in $\kreuzprodukt$ we  set $A_x:=\pi_x(A)$,
$A^{{\hat t}}:=\pi^{{\hat t}}(A)$ and $\widehat{A}:=U A U^\ast$.  For $A
\in \nkreuzprodukt$ we define the $A_x$ and $A^{{\hat t}}$ by
$$A=\int_X^\oplus A_x\,dm\,\;\hbox{and} \,\; U A
U^\ast=\int_{\widehat{G}}^\oplus A^{{\hat t}}\, d {\hat t}.$$ 
The fact that these families are only defined up to a set of measure
zero will be no inconvenience.

\rem{1} It is always possible to choose the $A_x$ such that
$$ T_t ^\ast A_{\alpha_t(x)} T_t=A_x$$
holds for all $x\in X$ and all $t\in G$. This can be seen in the
following way:
Theorem 2.1.7 implies
$\nkreuzprodukt\subset\{T_t\otimes S_{-t}\,|\,t\in G\}'.$
In particular, we have for fixed  $t\in G$
$$T_t  A_{\alpha_t(x)} T_t^\ast=A_x,\;a.e.\,x\in X.$$
Therefore we get, using the Fubini Theorem, that the family of operators
defined by 
$$\skp{\widetilde{A_x}\xi}{\eta}:=M( t\mapsto \skp{T_t
  A_{\alpha_t{x}}T_t^\ast \xi}{\eta}),$$ 
where $M$ is the mean on the abelian group $G$,  coincides almost
everywhere with the family $A_x$. Moreover, it is easy to see that the
family $\widetilde{A_x}$ has the required invariance property. 
\endrem

\subhead{2.2 The trace on  $\nkreuzprodukt$}\endsubhead

In the last section we proved that $\nkreuzprodukt$ is generated by a Hilbert algebra. This  allows to introduce a canonical trace on $\nkreuzprodukt$. 

We start with a simple lemma that will allow us to prove the equality of certain weights by proving the equality of the restrictions of these weights to suitable sets.

\lem{2.2.1} Let ${\Cal J}$ be a strongly dense ideal in a von Neumann algebra ${\Cal N}\subset B(H)$ containing the identity $I$ of $B(H)$.

(a) There is an increasing net $I_\lambda$ in ${\Cal J}$ converging strongly towards $I$. If $H$ is separable, $(I_\lambda)$ can be chosen as a sequence.

(b) If $\tau_1$ and $\tau_2$ are normal weights on  ${\Cal N}$ with $\tau_1(A A^\ast)=\tau_2(A A^\ast)<\infty$ for $A\in {\Cal J}$, then $\tau_1=\tau_2$ on $({\Cal J}{\Cal J})^+$.

(c) If $\tau_1$ and $\tau_2$ are normal weights on ${\Cal N}$, whose restrictions to $({\Cal J} {\Cal J })^+$  coincide, then $\tau_1=\tau_2$.
\endlem

\proof{}(a) The Ideal  ${\Cal J} {\Cal J }$ is normdense in the $C^\ast$ algebra ${\Cal C}:=\overline{{\Cal J}{\Cal J }}$. By 1.7.2 in \cite{17}, there exists therefore an approximate unit $I_\lambda$ in $ {\Cal J}{\Cal J }$    for ${\Cal C}$. As the net $I_\lambda$ is bounded and  increasing, it converges strongly to  some $E\in B(H)$ with
$$ E C= C E = C,\;\, C\in {\Cal C}.$$
As ${\Cal J}$ is strongly dense in $\Cal N$, the algebra ${\Cal C}$ is weakly dense in ${\Cal N}$  and 
$ E C= C E = C,\;\, C\in {\Cal N},$
follows.  This implies $E=I$. If $H$ is separable it is possible to choose an increasing subsequence of $(I_\lambda)$ converging to $E$.

(b) This follows using polarisation.

(c) Let $(I_\lambda)$ be as in (a). Fix $A=C C^\ast $ in ${\Cal N}^+$. As $\tau_1$ and $\tau_2$ are normal, it is enough to show
$$\tau_1(C I_\lambda C^\ast)=\tau_2(C I_\lambda C^\ast). $$
But this is clear, as  ${\Cal J} {\Cal J }$ is an ideal and as $(I_\lambda)$ belongs to  $({\Cal J} {\Cal J })^+$. \hfill $\qed$\endproof


\theo{2.2.2} There exists a unique normal trace $\tau$ on $\nkreuzprodukt$ with
$$ \tau(A A^\ast) =\skp{a}{a}$$
for $A$ with kernel $a$.
The trace $\tau$ is semifinite and normal and $\tau = \tau_{c} \circ Ad_{W^\ast}$, where $\tau_c$ is the canonical trace on ${\Cal L}({\Cal A})$ (cf. 2, ch 6, I of \cite{18}). Moreover
$$(\nkreuzprodukt)_{\tau}^2:=\{A\in \nkreuzprodukt\,:\,\tau(A A^\ast)<\infty\}=\kt.$$
\endtheo
\proof{} Clearly the identity of  $B(\lzgx)$ is contained in    $\nkreuzprodukt=\pi(\kreuzprodukt)''$  and we can apply the foregoing lemma with ${\Cal J}=\kt$ to get the uniqueness. 

As $Ad_W$ is an isomorphism by Theorem 2.1.4.,  the remaining statements follow easily from the corresponding statements  in 2, ch 6, I of \cite{18}.\hfill $\qed$
\endproof

\defi{2.2.3} For a selfadjoint operator $A$  affiliated to the von Neumann algebra  $\nkreuzprodukt$, i.e. whose resolution of the identity, $E_A$ \cite{35},  is contained in $\nkreuzprodukt$, define
$$\mu_A(B):=\tau( E_A(B))$$
for Borel measurable $B\subset R$. The map $\mu$ is called the {\it integrated  density of states} (IDS) for $A$ (cf. \cite{6}).
\enddefi

We mention that there is a  different approach to the IDS for one dimensional Schr\"odinger operators due to Johnson and Moser (cf. \cite{16, 24} ).

From Theorem 2.2.2 we get the following corollaries.

\cor{2.2.4} Let $A$ and $\mu_A$ be as in the preceding definition. Then $\mu_A$ is a spectral measure for $A$.
\endcor
\proof{} This is clear, as $\tau$ is faithful and normal. \hfill $\qed$ \endproof    

\cor{2.2.5} Let $A$ and $\mu$ be as in Corollary 2.2.4. If
  there exists a set $\sigma\subset \hbox{R}$ with $\sigma(A_x)=\sigma$
  a.e. $x\in X$, then $\sigma=supp\, \mu$.
\endcor

\proof{} As $\mu$ is a spectral measure of $A$, we have
$\sigma(A)=supp\, \mu$. By $\sigma(A_x)=\sigma$ a.e. $x\in X$, the
equality $\sigma=\sigma(A)$ holds.  \hfill $\qed$ \endproof      

There is another way to calculate the trace that  can be seen as an  application of \cite{13} (cf. Remark 1  below).

\lem{2.2.6} Let $A$ be in $(\nkreuzprodukt)_+$. Then there exists a unique $\Lambda(A)\in [0,\infty]$ with
$$\Lambda(A)\int_G g^2(t) dt=\int_X \tr(M_g A_x M_g) \,dm$$
for positive $g\in L^\infty(G)$, where $\tr$ denotes the usual trace on $B(\lzgx)$.
\endlem

\proof{} Uniqueness is obvious. Existence will follow from the uniqueness of the Haar measure on G, once we have shown that the RHS of the equation induces an invariant measure.

As $A$ is positive there exists $C\in \nkreuzprodukt$ with $A=C^\ast C$. We calculate
$$\mu(B):=\int_X  \tr(M_{\chi_B} A_x M_{\chi_B})dm =\int_X  \tr(M_{\chi_B} C^\ast_x C_x M_{\chi_B})dm.$$
Using that $\tr$ is a trace we conclude
$\mu(B) = \int_X \tr(C_x M_{\chi_B} C_x^\ast)dm.$

This formula and some simple monotone convergence arguments show that $\mu$ is a measure with 
$$\int_G g^2 d\mu=\int_X \tr(C_x M_{g^2} C_x^\ast)dm= \int_X \tr(M_g A_x M_g) \,dm.$$
It remains to show that $\mu$ is translation invariant. As $\nkreuzprodukt$ is contained in $\{T_t\otimes S_{-t}\,:\,t\in G\}'$ for each $t\in G$, the equation
$$C_{\alpha_t(x)}= T_t^\ast C_x T_t$$
holds for a.e. $x\in X$. This allows to calculate
$$\eqalign{ \mu(B-t)&=\int_X\tr( C_x M_{\chi_{B-t}} C_x^\ast) dm\cr
&=\int_X \tr( C_x T_t M_{\chi_{B}}T_t^\ast C_x^\ast) dm\cr
(\tr \;\hbox{is trace })\;&=\int_X \tr(T_t^\ast C_x T_t M_{\chi_{B}}T_t^\ast C_x^\ast T_t) dm\cr
&=\int_X\tr( C_{\alpha_t(x)} M_{\chi_{B}} C_{\alpha_t(x)}^\ast) dm\cr
&=\mu(B).\cr}
$$
Here we used in the last equation that $m$ is translation invariant. The calculation  shows that $\mu$ is translation invariant. This finishes the proof.\hfill $\qed$
\endproof

\theo{2.2.7} $\Lambda=\tau$.\endtheo
\proof{} As $\kt$ is a strongly dense ideal in $\nkreuzprodukt$ by  Proposition 2.1.6.,  it is by Lemma 2.2.1 enough  to show
$$\Lambda(A^\ast A)=\tau (A^\ast A)$$
for $A\in \kt$. Choosing a positive $g\in L^\infty (G)$ with $\int_G g^2 dt=1,$ we calculate for $A\in \kt$ with kernel $a\in \lzgx$
$$\eqalign{\Lambda(A^\ast A)&=\int_X\tr(M_g A_x^\ast A_x M_g)\,dm\cr
&=\int_X \int_{G\times G} g(t) a(t-s,\alpha_t(x))|^2 dt\, ds\, dm\cr
(Fubini)\;&=\int_G |g(t)|^2(\int_G \int_X|a(t-s,\alpha_t(x))|^2\,dm \,ds )\,dt \cr
(m,ds\;\hbox{transl. inv. })&=\int_G|g(t)|^2(\int_G\int_X |a(s,x)|^2 \,dm \,ds)\, dt\cr
(\|g\|_{L^2(G)}=1)\;&=\skp{a}{a},\cr}
$$
where we used that for an operator $K\in B( \lzg)$ with kernel $k\in L^2(G\times G)$ the eqality $\tr (K K^\ast)=\int_{G\times G} |k(t,s)|^2 dt ds $ holds.
\hfill $\qed$
\endproof

In some cases (e.g. in the almost periodic case or if $G=R^n, Z^n$) it is known that there exists a sequence $H_n\subset G$ with
$$ \lim_{n\to\infty} \frac{1}{m_G(H_n)} \int_G \chi_{H_n}(s) f(\alpha_s(x)) dm_G(s)=\int_X f(z) \,dm(z)
$$
for $f\in L^1(X,m)$ and $x\in X_f$, where   $X_f$ is a suitable subset of $X$  with $\mu(X-X_f)=0$. 
 If this is valid, and if $A\in \kt$ has a kernel  $a$ s.t. 
$x\mapsto \int_G |a(t,x)|^2 dt$ belongs to $ L^1(X,m)  $, then
$$\lim_{n\to\infty} \frac{1}{m_G(H_n)}\tr(\chi_{H_n}A_x A_x^\ast \chi_{H_n})=\tau(A A^\ast), \;\,\;\,\;\,\;\,\;\, (*)  $$
by $ \tr(\chi_{H_n}A_x A_x^\ast \chi_{H_n})= \int_G \int_G \chi_{H_n}(t) |a(s,\alpha_{-t}(x))|^2 ds \,dt$.
Here, one can interpret a  term of the form  $\chi_H B_x \chi_H$ as the restriction $B_x|_H$ of  $B_x$ to $L^2(H ,m_G|_H)$. One is in particular interested in the case, where $A_x=\chi_I(B_x)$ belongs to the resolution of the identity of $B_x$. As  restrictions of operators are comparatively accessible, the question arises whether 
$$\lim_{n\to\infty}\frac{1}{m_G(H_n)}\left(\tr(\chi_{H_n}\chi_I(B_x)   \chi_{H_n}) -  \tr( \chi_I(B_x|_{H_n}))  \right)=0$$
 for $x\in X$ and  $I\subset R$. If this can be established, the  equation 
$$ \mu_A(I)=\lim_{n\to \infty}  \frac{1}{m_G(H_n)}\tr(\chi_I(B_x|_{H_n})) $$
follows from $(*)$ immediately, as the equation $\chi_I( C )=\chi_I(C ) \chi_I(C )^\ast$ holds for arbitrary operators $C$ .  Note that $  \tr( \chi_I(B_x|_{H_n})) $ is just the number of eigenvalues of $B_x|_{H_n}$ in $I$.  For $I=(-\infty,E]$, $E\in R$, this type of  equation has been established for pseudodifferential operators with almost periodic coefficients in \cite{36}, for Schr\"odinger operators in \cite{6} using heat equation methods and for discrete $G$  in \cite{3}. It is called {\it  Shubin's trace formula}.

\rems{1} In \cite{13} transverse functions and transverse measures on groupoids are introduced and studied (cf. I, 5 of \cite{14} as well). It is possible to give $G\times X$ the structure of  a groupoid. The measure $m$ then induces  a unique transversal measure $\Lambda$ with certain properties. It is possible to show that $\Lambda$ satisfies the equation
$$ \Lambda (\xi) \int_G f\,dt =\int_X \xi_x(f)\,dm$$
for transverse functions $\xi$ and $f\in L^\infty(G)$. A direct calculation shows that for $A\in (\nkreuzprodukt)_+$ the mapping $\xi^x(B):=\tr(\chi_B A_x \chi_B)$ is a transverse function (if the components $A_x$ are chosen according to Remark 1 in Section 2.1). In this context Theorem 2.1.7 says essentially
$\Lambda(\xi_A)=\tau(A)$.

2. In \cite{1}  it
is shown that for a family $A_\omega$, $\omega \in \Omega$, of almost
periodic Schr\"odinger operators and $F\in C_0(\hbox{R})$ the equation 
$$\int_\Omega \tr(M_g F(A_\omega) M_g)dm_\Omega =\int_{\hbox{R}} F dk$$
 holds, where the measure $dk$ is given by a certain limit procedure. Using Definition 2.2.3 and Theorem 2.2.7 we see $dk = d\mu_A$.
\endrem

\subhead{2.3 Some special cases   }\endsubhead

If $m(X)<\infty$ (e.g. if $X$ is compact) there exist two alternative formulas for the trace on $\nkreuzprodukt$. They will now be discussed.

Define for  $A\in \nkreuzprodukt$ the operator $A_m:\lzg\longrightarrow \lzg$ 
by
$$\skp{A_m\xi}{\eta}:=\int_X\skp{A_x\xi}{\eta} dm, \;\, \xi, \eta \in
\lzg. $$

Since  $A_{\alpha_t{x}}= T_t A_x T_t^\ast$ a.e. $x\in X$ for fixed $t\in G$
and  $m$ is invariant under $\alpha$, the operator $A_m$ is
translation invariant. Therefore there exists  $\varphi \in
L^\infty(\widehat{G})$ with $A_m=F^{-1} M_\varphi F$, where $M_\varphi$
denotes the operator of multiplication by $\varphi$. Now it is easy to see
that the mapping 
$$J:\nkreuzprodukt\longrightarrow M(L^\infty(\widehat{G})),\;A\mapsto F A_m
F^{-1}$$ is linear, positive and faithful on 
$(\nkreuzprodukt)_+$. Let $\tau_\infty$ be the usual trace on $
M(L^\infty(\widehat{G}))$, i.e.  $\tau_\infty (\varphi):=\int_{\widehat{G}}
\varphi\, d{\hat t}.$

Moreover, define $\mu :(\nkreuzprodukt)_+\longrightarrow [0,\infty]$ by
$$\mu(A):=\int_{\widehat{G}} \skp{A^{{\hat t}}1}{1}\, d{\hat t},$$
 where $1$ denotes the function of $\lzdg$ with constant value 1. 
Then the following holds.

\theo{2.3.1} $\tau=\mu=\tau_\infty \circ J$.\endtheo
\proof{} We will show (1) $\tau=\tau_\infty \circ J$ and (2) $\tau= \mu$.

(1) $\tau=\tau_\infty \circ J$: By Theorem 2.2.7, it is enough to show $\Lambda(A)=\tau_\infty \circ J(A)$ for $A\in(\nkreuzprodukt)_+$.  For such an $A$ let $A_m$ and the function $\phi$ be  defined as above, i.e. $A_m=F^{-1} M_\phi F$ and $ M_\phi=J(A)$.  Choosing a positive $g\in L^{\infty}(G)$ with $\|g\|_{\lzg}=1$ we calculate
$$\Lambda(A)=\int_X \tr(M_g A_x M_g) dm=\tr (M_g A_m M_g)=\tr(M_g F^{-1} M_{\phi^{1/2}} F F^{-1} M_{\phi^{1/2}} F M_g).$$

For $\phi\in L^1(\widehat{G})$ the operator $K=M_g F^{-1} M_{\phi^{1/2}} F$ is a Hilbert Schmidt operator with kernel $k(t,x)=g(t) F^{-1}(\phi^{1/2})(t-s)$.  Thus the formula  $\tr (K^\ast K)=\int_{G\times G} |k(t,s)|^2 dt ds$ holds and  we get
$$\Lambda(A)=\int_G \int_G |g(t)  F^{-1}(\phi^{1/2})(t-s)|^2 dt ds=
\|\phi^{1/2}\|_{L^2(\gd)}=\int_{\gd} \phi(\hat{t}) d\hat{t}, $$
where we used the translation invariance of dt, $\|g\|_{L^2(G)}=1$, and the fact that the Fourier transform is an isometry. 

For arbitrary $\phi$ the equality $\Lambda(A)=\int_{\gd} \phi(\hat{t}) d\hat{t}$ now follows by a simple monotone limit procedure. 

(2) $\tau = \mu$: For $A\in (\nkreuzprodukt)_+$ let  $A_m$ and $\varphi$ be  as
above. By (1), it is enough to show $\varphi({\hat t})=
\skp{A^{{\hat t}}1}{1}=:\psi({\hat t})$ for a.e. ${\hat t}\in\widehat{G}$. But this
follows from the following calculation valid for all $\eta, \xi\in
\lzdg$:
$$\eqalign{
\skp{M_\varphi\xi}{\eta}&=\skp{A_m F^{-1} \xi}{F^{-1}\eta}\cr
&= \int_X \skp{A_x F^{-1} \xi}{F^{-1}\eta} dm\cr
&=\skp{A (I\otimes F^{-1})(1\otimes \xi)}{(I\otimes F^{-1})(1\otimes\eta)}\cr
&= \skp{A  U^\ast (1\otimes \xi)}{U^\ast (1 \otimes\eta)}\cr
&= \skp{\hat{A} (1\otimes \xi)}{(1\otimes\eta)}\cr
&= \skp{M_\psi \xi}{\eta}\cr}
$$
The theorem is proven. \hfill $\qed$ 
\endproof

\rems{1} The expression $\mu$ was used in \cite{25} (cf. also \cite{9}).

2. The mapping $J$ was first introduced by Coburn, Moyer and Singer  \cite{10} (cf. also \cite{36}) in  their paper on almost periodic operators.

3. If $X$ consists of only one point it is a forteriori
compact.
The positive operators in $\kreuzprodukt$ are just the
operators  $A=F_G^{-1} M_\varphi F_G$ where $\varphi\in C_0(\widehat{G})$ is
positive (cf. Remark 1 in Section 1).
The trace of such an  $A$ is then given by
$$\tau(A)=\int_{\widehat{G}} \varphi({\hat t})d{\hat t},$$
in particular $\tau(A)$ is  finite  iff $\varphi$ belongs to
$L^1(\widehat{G})$.
As the ideal of trace class operators  consists of the finite linear
combinations of positive operators with finite trace, we conclude that
$A=F_G^{-1} M_\varphi F_G$ is trace class iff $\varphi$ belongs to
$L^1(\widehat{G})\cap  C_0(\widehat{G})  $. 

In particular it is not true in general that an operator of the form
$A=\pi(\varphi)=F^{-1} M_{F(\varphi)} F$ with $\varphi \in \ccgx=C_c(G)$ is trace class. This shows that it is not possible (as  is sometimes done) to define a trace on $\nkreuzprodukt$ by setting
$$\tau (\pi(\psi)):=\int_X \psi(0,x) dm.$$
\endrem
We close this section with some remarks on the special case that $G$ is discrete. As in this case the function $\delta_e:G\longrightarrow C$ defined by $\delta_e(t)=1$ for $t=e$ and $\delta_e(t)=0$ for $t\neq e$ is positive, bounded with $\|\delta_e\|_{\lzg}=1$, we get easily
$$\tau(A)=\int_X \skp{A_x\delta_e}{\delta_e} \,dm.$$
Moreover, it is possible to show that there exists a  conditional expectation 
$$J:\nkreuzprodukt\longrightarrow L^{\infty}(X)$$
with $\tau=\tau_{\infty}\circ J$, where $\tau_{\infty}$ is the usual trace on $L^{\infty}(X)$ (cf. \cite{31} for a thorough discussion of this case). 

\head{3. Groups acting on groups}\endhead
This section is devoted to the study of a group acting on another group. This situation arises in particular in the context of almost periodic functions (cf.  \cite{8, 30}) and this is indeed the motivating example. The irrational rotation algebras, which have recieved a lot of attention (cf. \cite{15, 34} and references therein), arise in such a situation. They are used in the treatment of the so called Harper's model \cite{4}. The corresponding operator is  just the    almost Mathieu operator  for $\lambda=2$ (cf. \cite{22} and references therein for details about the almost Mathieu operator).

\subhead{3.1 The general case }\endsubhead

We will look at the following situation: Let $G$ and $X$ be locally
compact abelian groups, and let $j:G\longrightarrow X$ be a group
homomorphism. This  induces a homomorphism
$j^\ast:\xd \longrightarrow \gd $, where $\xd $ (resp.
$\gd $) denotes the dual group of $X$ (resp. $G$). Then there is an
action $\alpha$  of $G$ on $X$ given by 
$$\alpha_t:X\longrightarrow X,\;\, \alpha_t(x):=x+ j(t),$$
and an action of $\xd$ on $\gd$ given by
$$\ad_{\hat{x}}:\gd\longrightarrow \gd,\;
\ad_{\hat{x}}({\hat t}):={\hat t} +j^\ast(\hat{x}).$$ 
Similarly to the   unitaries $T_t$ and $S_t$ resp.,  acting on $\lzg$
and $\lzx$ resp. for $t\in G$,  there are unitaries $T_{\hat{x}}$ and
$S_{\hat{x}}$ defined by
$$\eqalign{ S_{\hat{x}}:\lzdg\longrightarrow\lzdg & \;\;
  S_{\hat{x}}\xi({\hat t}):=\xi(\ad_{(-\hat{x})}({\hat t})),\cr
T_{\hat{x}}:\lzdx\longrightarrow\lzdx & \;\;
  S_{\hat{x}}\xi(\hat{y}):=\xi(\hat{y}-\hat{x}).\cr}
$$

Moreover we define $$\ww:\lzdgdx\longrightarrow \lzdgdx\,\;\hbox{ by} 
\,\; \ww \xi(\hat{x},{\hat t}):=\xi(\hat{x},\ad_{\hat{x}}({\hat t}))$$ and
$\widetilde{U}$ by 
$$\widetilde{U}:=(F_X\otimes F_G) W^\ast = (F_X \otimes I) U,$$
where $F_X$ and $F_G$ are the Fourier transfom on $\lzx$ and $\lzg$
resp.  and $U$ and $W$ are as defined in the first section.

We will first  establish a spatial isomorphism between the von Neumann
algebras $\nkreuzprodukt$ and $\ndkreuzprodukt$. This can be done quite easily,
as the generators of these von Neumann algebras are  known explicitly. 

We will then provide proofs for some statements first appearing in
\cite{6} and \cite{7}, that yield much more, namely an isomorphism
between $\kreuzprodukt$ and $\dkreuzprodukt$.

We will need the following propositions.

\prop{3.1.1} (a) $W^\ast (T_t\otimes I)= (T_t\otimes S_t )
  W^\ast$,   $\ww^\ast (T_{\hat{x}}\otimes I)=(T_{\hat{x}}\otimes
  S_{\hat{x}}) W^\ast.$

(b) $ W(T_t \otimes I)= (T_t \otimes S_{-t}) W$,   $\ww (T_{\hat{x}}\otimes
I)= (T_{\hat{x}}\otimes S_{-\hat{x}}) W$.

(c) $\widetilde{U} W = \ww^\ast \widetilde{U}$. 

\endprop

\proof{} This can be seen by  direct calculation.\hfill $\qed$ \endproof    

\prop{3.1.2} (a) The von Neumann algebra  $\nkreuzprodukt$ is generated by
  operators of the form $T_t\otimes I$, $t\in G$, and the  operators of
  multiplication with the functions
$$ (t,x)\mapsto
\dk{\hat{x}}{\alpha_t(x)}=\dk{\hat{x}}{x}\dk{\hat{x}}{j(t)},\;\,
\hat{x}\in \hat{X}.$$
 (b) The von Neumann algebra $\ndkreuzprodukt$ is generated by operators of the form
$T_{\hat{x}}\otimes I$, $t\in G$, and the operators of multiplication
with the functions
$$ (\hat{x},{\hat t})\mapsto
\dk{t}{\ad_{\hat{x}}({\hat t})}=\dk{t}{j^\ast(x)}\dk{t}{{\hat t}}$$
\endprop

\proof{} This follows from Theorem 2.1.7  and the well
known fact that the characters generate the von Neumann algebra
$L^\infty$.\hfill $\qed$ \endproof

Now we can  prove the spatial isomorphism, mentioned above.

\theo{3.1.3} The unitary $\widetilde{U}: \lzgx\longrightarrow \lzdgdx$
  establishes a spatial isomorphism between $\nkreuzprodukt$ and
  $\ndkreuzprodukt$.
\endtheo
\proof{} By the foregoing proposition, it is enough to show
{\parindent=1cm
\item{(1)}$ \widetilde{U}( T_t\otimes I) \widetilde{U}^\ast =M( \dk{t}{j^\ast(x)}\dk{t}{{\hat t}}  ),$
 \item{(2)}$ \widetilde{U} M( \dk{\hat{x}}{x}\dk{\hat{x}}{j(t)}   ) \widetilde{U}^\ast = T_{\hat{x}}\otimes I$,
\par}
where $M(\varphi)$ denotes as usual the operator of multiplication by
$\varphi$.

(1) Using Proposition 3.1, we can calculate
$$\eqalign{ \widetilde{U} (T_t\otimes I)&= (F_X\otimes F_G)
  W^\ast (T_t\otimes I)\cr
&=  (F_X\otimes F_G) (T_t\otimes S_t) W^\ast\cr
&=\dk{t}{\cdot}\dk{t}{j^\ast\cdot} (F_X\otimes F_G)  W^\ast. \cr}
$$

(2) This can be seen by similar arguments. \hfill $\qed$ \endproof    

As already stated above there is even an isomorphism between the $C^*$-algebras  $\kreuzprodukt$ and
$\dkreuzprodukt$. We will establish this isomorphism in two steps. In
the first step we will show that the image of an  operator with $L^2$-kernel under conjugation by $\widetilde{U}$  is also an operator with an
$L^2 $  kernel. In the second step we will prove that conjugation by
$\widetilde{U}$   is  actually an isomorphism between $\kreuzprodukt$ and
$\dkreuzprodukt$. We remark that both of these facts have already been
stated in \cite{6} and \cite{7}, where, however, no proof was given.

\lem{3.1.4}
 Let $a\in \lzgx$ be the kernel of a bounded operator $A$ on
  $\lzgx$, i.e. 
$$ A\xi(t,x)=\int_G a(t-s,\alpha_s(x)) \xi(s,x) ds.$$
Then $\widetilde{A}:=\widetilde{U} A\widetilde{U}^\ast $ is a bounded operator on
$\lzdgdx$ with kernel
 $$\hat{a}:=(F_X\otimes F_G) W a\in \lzdgdx,$$
 i.e.
$$\widetilde{A} \xi(\hat{x},{\hat t})=\int_{\hat{X}}
\hat{a}(\hat{x}-\hat{y},\ad_{\hat{x}}({\hat t})) \xi(\hat{y},{\hat t})
d\hat{y}.$$
\endlem

\proof{} Let $\hat{\xi}:=(F_X\otimes F_G) (\xi)$ be an arbitrary
function in $\lzdgdx$. We calculate

$$\eqalign{
\widetilde{U} A \widetilde{U}^\ast \hat{\xi}(\hat{x},{\hat t})&= (F_X\otimes
F_G) W^\ast  A W \xi (\hat{x},{\hat t})\cr
&=(F_X\otimes F_G)((t,x)\mapsto \int_G a(t-s,x)
W\xi(s,\alpha_{-t}(x))ds)(\hat{x},{\hat t})\cr
= (F_X\otimes F_G)(&(t,x)\mapsto \int_G (W a)(t-s,\alpha_{t-s}(x))
\xi(s,\alpha_{t-s}(x)) ds)(\hat{x},{\hat t}). \cr}
$$
As for fixed $t\in G$ the mapping $x\mapsto \int_G| (W a)(t-s,\alpha_{t-s}(x))
\xi(s,\alpha_{t-s}(x))| ds$ belongs to  $L^1(X)$, this expression equals
$$
 (I\otimes F_G)\left(\int_X \dk{\hat{x}}{-x}\left(\int_G (W a)(t-s,\alpha_{t-s}(x))
\xi(s,\alpha_{t-s}(x))ds\right) dx\right)(\hat{x},{\hat t}),$$
which yields after the substitution $(x\mapsto x-j(t-s))$
$$\eqalign{
. . .&=(F_X\otimes F_G)\left((t,x)\mapsto 
  (\dk{\hat{x}}{-j(\cdot)}Wa)(\cdot,x)\ast
  \xi(\cdot,x)(t)\right)(\hat{x},{\hat t})\cr
&=F_X\left(x\mapsto (I\otimes F_G)(Wa)({\hat t} +j^\ast( \hat{x}),x) (I\otimes
  F_G)(\xi)({\hat t},x)\right)(\hat{x})\cr
&= (F_X\otimes F_G)(Wa)({\hat t}+j^{\ast}( \hat{x}),\cdot)\ast(F_X\otimes
F_G)(\xi)({\hat t},\cdot)(\hat{x})\cr
&= \int_{X} \hat{a}(\hat{x}-\hat{y},\ad_{\hat{x}}({\hat t}))
\hat{\xi}(\hat{y},{\hat t})d\hat{y}.\cr}
$$
This proves the lemma.\hfill $\qed$ \endproof    

\theo{3.1.5} The mapping $Ad_{\widetilde{U}}:
  \kreuzprodukt\longrightarrow \dkreuzprodukt,\,\; A\mapsto \widetilde{U}
  A\widetilde{U}^\ast$ is an isomorphism of $C^\ast$-algebras.
\endtheo

\proof{}  We have  to show that $Ad_{\widetilde{U}}(\kreuzprodukt)$
is  contained in 
$\dkreuzprodukt$ and that   $Ad_{\widetilde{U}^\ast}(\dkreuzprodukt)$ is
 a subset of $\kreuzprodukt$.  We show 
{\parindent=1cm
\item{(1)} There is a dense  set  $D\subset \kreuzprodukt$ with
  $Ad_{\widetilde{U}}(D)\subset \dkreuzprodukt$.
\item{(2)} There is a dense subset $F$ of $\dkreuzprodukt$ with
  $Ad_{\widetilde{U}^\ast}(F) \subset \kreuzprodukt$. 
\par}

(1) Let $$D:=\{\pi( W^\ast(g\otimes (h_1 \ast h_2)))\,|\, g\in
C_c(G),\,h_1,h_2\in C_c(X)\}.$$
It is easy to see that $D$ is in fact dense in $\kreuzprodukt$. The
lemma yields that for $A=\pi( W^\ast(g\otimes (h_1 \ast h_2)))\in D$ the
operator $Ad_{\widetilde{U}}(A)$ has the kernel 
$$\hat{a}:= F_G(g)\otimes (F_X(h_1) F_X(h_2)).$$
 But, as $F_G(g)$ belongs to $C_0(\widehat{G})$ and
$(F_X(h_1) F_X(h_2))$ belongs to $L^1(\widehat{X})$, the function
$\hat{a}$ is indeed the kernel of an operator in $\kreuzprodukt$. 

(2) This can be seen by similar arguments. \hfill $\qed$ \endproof    

We now provide a proof for  another  theorem which was already stated (without proof) in \cite{6} and \cite{7}.

\theo{3.1.6}
 Let $\tau$  (resp.  $\hat{\tau}$)  be the trace on
  $\nkreuzprodukt$  (resp.  $\ndkreuzprodukt$) defined in the last
  section. Then  the equation 
$$ \tau(A)= \hat{\tau}( \widetilde{U} A \widetilde{U}^\ast)$$
holds for all $A\in (\nkreuzprodukt)_+$.
\endtheo 

\proof{} It is enough to consider the case $A=B B^\ast$ with $B\in
\kt$ with kernel $b$ (cf. Lemma 2.2.1). Then
the kernel of $\widetilde{U} B \widetilde{U}^\ast $ is given by $(F_G\otimes
F_X)(Wb)$ and we have
$$\eqalign{ \hat{\tau}( \widetilde{U} A \widetilde{U}^\ast)&=
  \int_{\widehat{G}}\int_{\widehat{X}} |(F_G\otimes  F_X)(Wb)(\hat{x},{\hat t})|^2 d\hat{x} d{\hat t}\cr
&= \int_{G}\int_{X} |Wb(t,x)|^2 dt dx \cr
&= \int_X \int_G |b(t,x)|^2 dt dx \cr
&= \tau(B B^\ast). \cr
}
$$
As $A=B B^*$ the theorem is proven. \hfill $\qed$ \endproof    

There is an analogue of the classical Plancherel Theorem.

\cor{3.1.7}The mapping  $Ad_{\tilde{U}}$ establishes an isomorphism between the ideals  $(\nkreuzprodukt)_\tau^2$ and $(\ndkreuzprodukt)_{\widehat{\tau}}^2$ with 
$$\tau( A A^{\ast})= \widehat{\tau}(Ad_{\tilde{U}}(A) Ad_{\tilde{U}}(A)^{\ast}).$$                               \endcor
\proof{} This follows directly from Theorem 3.1.6. \hfill $\qed$ \endproof

We will now give a short  application of the above theory.

\subhead{3.2 Periodic operators}\endsubhead

Let $H$ be a closed subgroup of a locally compact abelian group $G$ such
that $X:=G/H$ is compact. Let $p:G\longrightarrow X$ be the canonical
projection. Then the machinery developped in the last section can be
applied with $j=p$. 
For $A\in \nkreuzprodukt$ we denote in this section by $\widehat{G}\ni
{\hat t}\mapsto A^{{\hat t}}$ the family of operators with
$$\int_{\widehat{G}}^{\oplus} A^{{\hat t}} d{\hat t} = \widetilde{U} A \widetilde{U}^{\ast},$$ 
where $\widetilde{U}$ was defined in the last section. 

The operators in $\nkreuzprodukt$ have a very strong
invariance property. 

\prop{3.2.1}
 For every $A\in \nkreuzprodukt$ there exist unique
  $A_x$, $x\in X$, with
(i) $A=\int_X^\oplus A_x dm$ and  (ii) $T_t A_{\alpha_t(x)} T_t^{\ast} =
A_x$, $ x\in X,t\in G$.

The same holds for selfadjoint $A$ that are affiliated to $\nkreuzprodukt$.

\endprop

\proof{} For $A\in \nkreuzprodukt$ the existence of such $A_x$ has already been shown (cf.  Remark 1 in Section 2). The uniqueness follows as $p$ is surjective. \
For selfadjoint $A$ affiliated to $\nkreuzprodukt$ the uniqueness proof is unchanged. Existence follows by looking at $(A+ i)^{-1}$.
\hfill $\qed$ \endproof    

We have the following theorem. 

\theo{3.2.2} Let $A$ be selfadjoint and affiliated to $\nkreuzprodukt$ with resolution
 of identity $E_A$ and fibres $A_x$, $x\in X$,  chosen according to the preceding
  proposition. Then the measure $\mu$    defined in Definition 2.2.3 is a spectral measure for all
  $x\in X$.
\endtheo

\proof{} By Corollary 2.2.4, the measure  $\mu$ is a spectral measure for
$A$. As all $A_x$, $x\in X$,  are unitarily equivalent by 
Proposition 3.2.1,  the statement follows.\hfill $\qed$ \endproof   

\rems{1} Kaminker and Xia show in \cite{25} by the use of a spectral duality principle  that certain elliptic periodic operators have purely continuous spectra on the complement of the set of discontinuities of $\lambda \mapsto \tau(E_A((-\infty,\lambda]))$. Theorem 3.2.2 shows in particuar that this holds for arbitrary periodic operators for purely algebraic reasons.

2. For periodic Schr\"odinger operators it is possible to show that the spectrum is purely absolutely continuous  using some analyticity arguments (cf. \cite{33} and references therein).
\endrem

\def\gh{\widehat{G}/H^{\perp}}

We finish this section with a short discussion of another formula for
$\tau$. Let $H^{\perp}\subset \widehat{G}$ be the annihilator of $H$ and
let 
$q:\widehat{G}\longrightarrow \gh$ denote the canonical projection. For
$g= f\circ q$ define $\{g\}:=f$. 
Let for  $\rho=q({\hat t})$ the functional
$I_\rho:L^\infty({\widehat{G}})\longrightarrow \hbox{R}$ be defined by
$$I_\rho(f):=\int_{H^{\perp}} f({\hat t}+ h^{\perp})
dm_{H^\perp}=\sum_{h\in H^{\perp} } f({\hat t}+ h^{\perp})   ,$$
then the  desintegration formula
$$ \int_{\widehat{G}} f({\hat t}) d{\hat t}=\int_{\gh} I_\rho(f)
dm_{\gh}(\rho)$$
holds (cf. \cite{20}).

Let $A\in (\nkreuzprodukt)_+$ be given.
Identifying $\widehat{(G/H)}$ with $H^{\perp}$ and using 
$$ \tr A^{{\hat t}}=\sum_{h\in H^{\perp}} \skp{A^{{\hat t}}
  \delta_{h}}{\delta_{h}}=I_{q({\hat t})}( \hat{s}\mapsto
  \skp{A^{\hat{s}} \delta_e}{\delta_e}),$$
we calculate
$$\eqalign{ \tau(A) &= \int_{\widehat{G}} \skp{A^{{\hat t}}\delta_e}{\delta_e} d{\hat t}\cr
&= \int_{\gh} I_\rho( \hat{s}\mapsto \skp{A^{\hat{s}}\delta_e}{\delta_e}) dm_{\gh}(\rho)\cr
&=\int_{\gh}\{\hat{s}\mapsto \tr A^{\hat{s}}\}(\rho) dm_{\gh}(\rho).\cr}
$$

 The RHS of this equation is
 essentially the integrated density of states defined in ch.
 XIII of  \cite{33} for periodic operators. 

\head{4. Spectral duality}\endhead

By spectral duality we mean a relation between the spectral types of
$A_x$, $x\in X$, and the spectral types of $A^{{\hat t}}$, ${\hat t}\in
\gd $ of the form "If  $A^{{\hat t}}$ has pure point spectrum a.e. ${\hat t}\in \gd$, then  $A^{{\hat t}}$   has purely (absolutely)
  continuous spectrum a.e. $\,x\in
X$."  

Theorems of this form have been stated  in \cite{7, 9, 25}. We cite the theorem of \cite{25}.

\theo{4.1} Let $A\in \nkreuzprodukt$ be selfadjoint with purely
  continuous spectrum on a Borel set $E$ s.t. $A^{{\hat t}}$  has pure point
  spectrum on $E$ for almost all ${\hat t}\in \dg$, then $A^{{x}}$ has purely
  coninuous spectrum on $E$ for almost all $x\in X$.
\endtheo

In \cite{22} another form of duality is proven for the Almost
Mathieu Equation.  The method developped there can be carried over with
only small changes to give 

\theo{4.2} Let $(Z,\alpha,X,m)$ be a dynamical system. Let $A\in
  Z\times_\alpha L^\infty(X)$ be selfadjoint with spectral family $E_A$ s.t. $A_x$ has only pure
  point spectrum with simple eigenvalues for almost all $x\in X$.  Then
  $$\mu(B):=\tau(E_A(B))$$
is a spectral measure for $A^{\eta}$ for almost all $\eta\in \widehat{Z}=:S^1$.
\endtheo

\proof{} As $\mu$ is a spectral measure for $A$ by Corollary 2.2.4, it is enough to show that there are spectral measures
$\nu^{\eta} $ for $A^{\eta}$ not depending on $\eta$. This is
shown following \cite{22}.

By the same method as in \cite{22}, it can be shown that there exist
measurable functions
$$ N_j:X\longrightarrow Z\cup \{\infty\},\;\,\hbox{and}\;\,
\varphi_j^l:X\longrightarrow l^2(Z),\;j\in Z,l\in N,$$
s.t. the set 
$$\{\varphi_j^l(x)\,|\, j\in Z,\;l=1,...,N_j(x)\}$$
is an orthonormal basis of $l^2(Z)$ consisting of eigenvectors of
$A_x$ for almost all $x\in X$ and that the  $\varphi_j^l$ satisfy
$$ \varphi_j^l(x)=T_k \varphi_{j-k}^l(\alpha_k(x)) ,\,\, k\in Z.$$
In particular we have
$$\skp{\varphi_j^l(x)}{T_k\varphi_j^l(\alpha_k(x))}_{\lzx}=0,\;k\neq 0,$$
as $\varphi_j^l$ and $T_k\varphi^l_j(\alpha_k(x))=\varphi_{j+k}^l(x)$ are
different members of an ONB.

The simplicity of the eigenvalues is crucial to get this measurable
section of eigenfunctions.  We will now show (cf. \cite{22})

{\parindent=1cm
\item{(1)}Fix $\psi\in \lzx, \;F\in C_0(\hbox{R}),\,j\in Z,\;l\in N$. Let
  $\xi(z,x):=\psi(x) \varphi_j^l(x)(z)$,
  $\hat{\xi}_{\eta}(x):=U\xi(\eta,x)$ and
$$ \mu_\eta(F)
:=\skp{ \hat{\xi}_{\eta} }{ F(A^\eta) \hat{\xi}_{\eta} }_{\lzx}.$$
Then $\mu_\eta(F)$ is independent of $\eta$ a.e. $\eta$.
\item{(2)}$\mu_\eta$ does not depend on $\eta$ a.e. $\eta$.
\item{(3)} There exist $\nu_\eta$  not depending on $\eta$
s.t. $\nu_\eta$ is a spectral measure for $A^\eta$ for a.e. $\eta$. 
\par}

By the remarks at the beginning of the proof, the theorem follows from
(3).

{
(1) It is enough to show $0= \int_{S^1} \dk{\eta}{z}  \mu_\eta(F) d\eta$
for all $z\in Z$ with $z\neq 0$.
We calculate
$$
\eqalign{\int_{S^1} \dk{\eta}{z}  \mu_\eta(F) d\eta&=
  \skp{(I\otimes M_{z}) U\xi}{U F(A) \xi}\cr
&=\skp{U (T_z\otimes S_z)\xi}{U F(A) \xi}\cr
&=\skp{(T_z\otimes S_z)\xi}{F(A) \xi}.\cr}
$$ 
Using that $\varphi_j^l(x)$ is an eigenvector  corresponding to the eigenvalue $e^l_j(x)$ say, we get 
$$
\eqalign{... &=\int_X
  F(e^l_j(x))\psi(\alpha_z(x))\psi(x)\skp{T_z \varphi_j^l(\alpha_z(x))}{\varphi_j^l(x)}dm\cr
&=0,\cr}
$$
where we used the relation $ \skp{T_z \varphi_j^l(\alpha_z(x))}{\varphi_j^l(x)}=0$.}

(2) As $C_0(\hbox{R})$ is separable, this follows from (1).

(3) Let $\{\psi_l\} $ be an ONB of $\lzx$. Then $\{\psi_l\otimes
\varphi_j^l\}$ is an ONB in $\lzgx$ and, as $U$ is unitary, it follows that
the
$\xi_{l,j,m}:=U(\psi_l\otimes \varphi_j^l)$
form an ONB in $\lzdgx$. Thus the set
$\Cal{T}:=\{\xi_{l,j,m}(\eta,\cdot)\,|\, l,j,m\}$
is total in $\lzx$ for
almost all $\eta\in S^1$. (Notice that the set  $M_\varphi:=\{\eta\in
S^1\,|\,\varphi\perp \xi_{l,j,m}(\eta,\cdot)\,\forall l,j,m\}$ has measure
zero for each $\varphi\in \lzg$.)
Therefore the measures $\nu^\eta$ defined by 
$$\nu^\eta(B):=\sum_{l,j,n}\skp{\chi_B(A^\eta)\xi_{l,j,m}(\eta,\cdot)}{
  \xi_{l,j,m}(\eta,\cdot)}_{\lzg}$$
are spectral measures for almost all $\eta\in S^1$, which do not depend
on $\eta$ by (2). 
The theorem  follows. \hfill $\qed$ \endproof    

{\it Acknowledgments.} The author would like to thank P. Stollmann for many use\-ful discussions. Financial support from Studienstiftung des deutschen Vol\-kes (Doktorandenstipendium) is gratefully acknowledged.

{\flushpar
\eightpoint
\addr{Daniel Lenz}
\addr{J. W. Goethe Universit\"at}
\addr{60054 Frankfurt/Main}
\addr{Germany}
\addr{{\it email: dlenz\@math.uni-frankfurt.de}}
}

\enddocument